\documentclass[twocolumn]{aastex701} 

\received{March 17, 2026}
\revised{April 23, 2026}
\accepted{May 8, 2026}

\shorttitle{A Third Galaxy Missing Dark Matter}
\shortauthors{Keim et al.}

\begin{document}

\title{A Third Galaxy Missing Dark Matter along a Trail of Galaxies in the NGC 1052 Field}

\correspondingauthor{Michael A. Keim}
\email{michael.keim@yale.edu}

\author[0000-0002-7743-2501]{Michael A. Keim}
\affiliation{Department of Astronomy, Yale University, PO Box 208101, New Haven, CT 06520-8101, USA}
\affiliation{Dragonfly Focused Research Organization, 150 Washington Avenue, Santa Fe, NM 87501, USA}
\email{michael.keim@yale.edu}

\author[0000-0002-8282-9888]{Pieter van Dokkum}
\affiliation{Department of Astronomy, Yale University, PO Box 208101, New Haven, CT 06520-8101, USA}
\affiliation{Dragonfly Focused Research Organization, 150 Washington Avenue, Santa Fe, NM 87501, USA}
\email{fakeemail1@google.com} 

\author[0000-0002-5120-1684]{Zili Shen}
\affiliation{Department of Astronomy, Yale University, PO Box 208101, New Haven, CT 06520-8101, USA}
\affiliation{Dragonfly Focused Research Organization, 150 Washington Avenue, Santa Fe, NM 87501, USA}
\email{fakeemail2@google.com} 

\author[0000-0002-1841-2252]{Shany Danieli}
\affiliation{Department of Astrophysical Sciences, 4 Ivy Lane, Princeton University, Princeton, NJ 08544}
\affiliation{School of Physics and Astronomy, Tel Aviv University, Tel Aviv 69978, Israel}
\email{fakeemail3@google.com} 

\author[0000-0002-7075-9931]{Imad Pasha}
\affiliation{Dragonfly Focused Research Organization, 150 Washington Avenue, Santa Fe, NM 87501, USA}
\affiliation{Department of Astronomy, Yale University, PO Box 208101, New Haven, CT 06520-8101, USA}
\email{fakeemail4@google.com} 

\begin{abstract}
While most dwarf galaxies are strongly dark matter dominated, two remarkable objects in the NGC\,1052 field, DF2 and DF4, appear to lack dark matter. DF2 and DF4 were recently found to be part of a trail of low luminosity galaxies that follow a linear relation between their position and radial velocity. If the other galaxies on this trail formed together with DF2 and DF4, e.g., from gas that was separated from dark matter through a `bullet dwarf' collision, they may lack dark matter as well. Here we constrain the mass of DF9, the galaxy on the trail that most closely resembles DF2 and DF4. Using Keck/KCWI we find that DF9's stellar velocity dispersion is $6.5^{+3.9}_{-4.3}$ km\,s$^{-1}$. This is consistent with the $8.3^{+0.9}_{-1.4}$ km\,s$^{-1}$ dispersion that is expected from DF9's $1.4\times 10^8$ M$_\odot$ stellar mass alone, and we conclude that -- like DF2 and DF4 -- dark matter is not required to explain the kinematics of DF9. The dispersion is far below the $24\pm3$ km\,s$^{-1}$ expected if DF9 had a $1.4\times 10^{10}$ M$_\odot$ dark matter halo falling on the stellar mass--halo mass relation. As demonstrated in Appendix~\ref{Sec:Dist}, these results are not sensitive to the assumed distance of the trail. Our results are further evidence that the trail of low mass galaxies in the NGC\,1052 field formed together in a unique galaxy formation channel, and are consistent with the prediction of the bullet dwarf scenario that other trail galaxies should show the same lack of dark matter as DF2 and DF4. 
\end{abstract}


\section{Introduction} \label{Sec:Introduction}


In the standard paradigm of galaxy formation, galaxies form within larger dark matter halos, starting out as gas which cools inside the halos and forms stars \citep{2010gfe..book.....M}. Nearly all galaxies are thought to be gravitationally dominated by dark matter. The fraction of dark matter in a galaxy depends on its mass, with low mass dwarf galaxies being the most dark matter dominated due to their low escape velocities allowing for supernovae and other `feedback' mechanisms to effectively heat and eject gas. As a result, such galaxies generally have dark matter halos which are $>$100$\times$ more massive than their mass in stars \citep{2013ApJ...770...57B}.


The discovery of two faint galaxies lacking dark matter, NGC 1052-DF2 and NGC 1052-DF4 (DF2 and DF4 hereafter), challenged this picture \citep{2018Natur.555..629V,2019ApJ...874L...5V}. In addition to having total masses consistent with their stellar masses alone \citep{2018Natur.555..629V,2018ApJ...863L..15W,2019ApJ...874L...5V,2019ApJ...874L..12D,2019A&A...625A..76E,2019ApJ...877..133D,2022ApJ...935..160K,2023ApJ...957....6S,2025A&A...697A.145F}, these two galaxies both have extremely luminous globular clusters \citep{2018ApJ...856L..30V,2019ApJ...874L...5V,2021ApJ...909..179S} and are spatially extended, ultra-diffuse galaxies \citep{2015ApJ...798L..45V}. Taken together, these properties set DF2 and DF4 apart from all other galaxies that were previously known.


\citet{2022Natur.605..435V} later discovered that DF2 and DF4 are part of a tight,  statistically significant linear alignment of a dozen faint galaxies in the NGC\,1052 field. A linear trail so tight relative to its local environment is unique among wider diffuse object catalogs (J.\ An et al., in preparation; analysis of \citealt{2021ApJS..252...18T,2022ApJS..261...11Z}). \citet{2025ApJ...988..165K} further found that these trail galaxies are kinematically connected, with their velocities linearly increasing based on their position along the trail, following the trend predicted by the positions and velocities of DF2 and DF4. This strongly suggests that the other trail galaxies had a common origin, forming together with DF2 and DF4. Following this line of reasoning, it may be supposed that whatever caused DF2 and DF4's lack of dark matter would have caused the other galaxies on the trail to lack dark matter as well.

Notably, a similar alignment of galaxies had previously been seen in simulations \citep{2020ApJ...899...25S,2021ApJ...917L..15L} of a galaxy formation scenario proposed by \citet{2019MNRAS.488L..24S} in which dark matter-free galaxies arise from gas which is separated out from its original dark matter due to a high-speed collision. This dwarf galaxy scale analog of the Bullet Cluster \citep{2006ApJ...648L.109C} explains both DF2 and DF4's extremely luminous globular clusters \citep{2020ApJ...899...25S,2021ApJ...917L..15L} and large sizes \citep{2022MNRAS.510.3356T}, and the fact that the globular clusters are nearly monochromatic \citep{2022ApJ...940L...9V}. Follow-up simulations after the discovery of the trail have confirmed that the observed trail, including its relative radial velocities, relative distances, and ages, is consistent with joint formation following a single high-speed `bullet dwarf' collision \citep{2024ApJ...966...72L,2025ApJ...988..165K}. However, perhaps the strongest prediction of this scenario -- that the other galaxies on the trail besides DF2 and DF4 lack dark matter too, having formed from the same dark matter-free gas -- is yet to be tested.

\begin{figure*}
    \centering
    \includegraphics[width=0.85\textwidth]{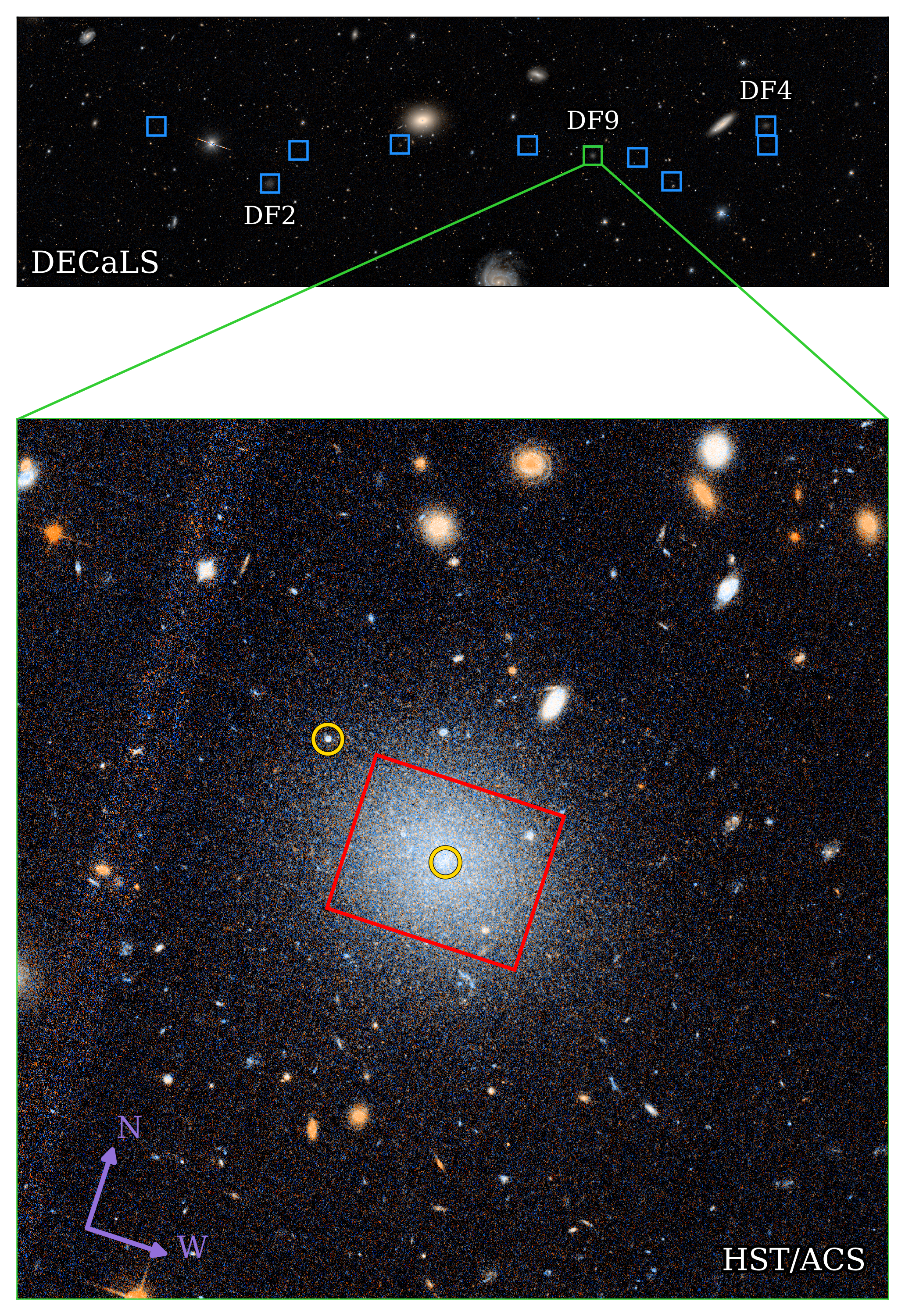}
    \caption{\textit{HST}/ACS image of DF9 shown below a DECaLS image of the NGC 1052 field. The kinematically-connected trail of galaxies, including DF2 and DF4, is indicated in blue boxes. The field-of-view for KCWI pointings is shown in red, and the two clusters with known radial velocities are circled in yellow. Both images are oriented along the trail axis \citep{2025ApJ...988..165K}. \label{Fig:Image}} 
\end{figure*}

The primary reason why the remaining galaxies' masses have not been measured is due to their faint, diffuse nature. DF2 and DF4 were discovered first as they are the brightest galaxies on the trail, though even DF2 and DF4 are `ultra-diffuse' \citep{2015ApJ...798L..45V}. The remaining galaxies are up to 100$\times$ fainter, making it difficult to measure their radial velocities (see \citealt{2025ApJ...988..165K}) and nearly impossible to measure their internal kinematics. Furthermore, the spectral resolution required to reliably measure the trail galaxies' low masses is even higher, given velocity dispersion follows the relation $\sigma^2 \propto M$. However, there is one trail galaxy for which it is feasible to measure a velocity dispersion: NGC\,1052-DF9 (DF9 hereafter). DF9, shown in Figure~\ref{Fig:Image}, stands out as the closest analog to DF2 and DF4, with a similar luminosity, size, and bright cluster population.\footnote{Indeed, DF9 is the closest trail galaxy to DF2 and DF4 in terms of globular clusters, however its brightest cluster is nuclear rather than globular for which extreme luminosity is not uncommon \citep{2023MNRAS.522..595B}. Even so, it is possible this formed from bright globular clusters which sank to the nucleus.} The expected dispersions from the stars and from the dark matter in DF9 are quite similar to those for DF2 and DF4 (see Section~\ref{Sec:ComparisontoExpectedDispersion}), making the techniques
used by \citet{2019ApJ...874L..12D} and \citet{2023ApJ...957....6S} to measure DF2 and DF4's dispersions with the Keck Cosmic Web Imager (KCWI) on Keck II ideal. 

The goal of this paper is therefore to measure the `dynamical' mass of DF9 through its velocity dispersion, in order to test the hypothesis that, like DF2 and DF4, the remaining galaxies on the trail lack dark matter. In Section~\ref{Sec:Data}, we describe how we obtain, reduce, and fit our data. In Section~\ref{Sec:Results}, we compare DF9's velocity dispersion to that expected from its stars and from a dark matter halo. We discuss the implications of our work in Section~\ref{Sec:Discussion} and present our conclusions in Section~\ref{Sec:Summary and Conclusion}.


\section{Data} \label{Sec:Data}


\subsection{Observations} \label{Sec:Observation}

Our KCWI observations were taken across 2 nights on 2024 October 2 and 3. We utilized the medium image slicer with the RH3 grating at a central wavelength of 8600 \r{A} and 2$\times$2 binning. Exposures were taken in pairs, with two successive 750 s on-target exposures at a time, to reduce the impact of cosmic rays. Given that the size of DF9 is larger than the field of view, for the purposes of sky subtraction we included equal length sky offsets pairs, i.e. two successive 750 s exposures, bracketing the on-target exposure pairs. The field of view for our on-target pointing is shown in Figure~\ref{Fig:Image}, as imaged using the \textit{Hubble Space Telescope (HST)} Advanced Camera for Surveys (ACS) in mid-cycle program 16912 (PI: van Dokkum). Its position relative to the rest of the trail is indicated in Dark Energy Camera (DECam) Legacy Survey (DECaLS; \citealt{2019AJ....157..168D}) imaging on the top panel. The sky offset is 1.2$\arcmin$ off to the south of the cutout shown. Conditions were good with no clouds, negligible moon, and good seeing ($0.5\arcsec-0.9\arcsec$). We obtained 24 on-target exposures and 28 sky exposures, for a total combined exposure time of 10.83 hrs. 


\subsection{Data Reduction} \label{Sec:DataReduction}


For initial frame reduction we utilized the KCWI Data Reduction Pipeline\footnote{\href{https://github.com/Keck-DataReductionPipelines/KCWI_DRP}{github.com/Keck-DataReductionPipelines/KCWI\_DRP}} (KCWI DRP). Rather than relying on its cosmic ray detection, we employed a custom technique adapted from the publicly available CRCombine package\footnote{\href{https://github.com/prappleizer/CRCombine}{github.com/prappleizer/CRCombine}} to identify cosmic rays by taking the difference between exposure pairs and masking affected and neighboring pixels that exceed a 99-percentile threshold in the noise-normalized difference image. To avoid artifacts associated with strong sky lines, we then conduct sky subtraction using the neighboring four sky frames, rescaled based on a weighting such that the strength of lines in the combined sky frame matches that of the sky lines in the 1D on-target spectra at wavelengths which are not affected by the galaxy's spectrum. Sky subtracted science frames, as well as sky frames, were then fed as inputs into the KCWI DRP (we note that products are overall similar to performing these steps after running the pipeline on raw frames; these efforts were made due to the KCWI DRP's treatment of strong sky lines which are more prevalent on the red side).

\begin{figure*}
    \centering
    \includegraphics[width=0.95\textwidth]{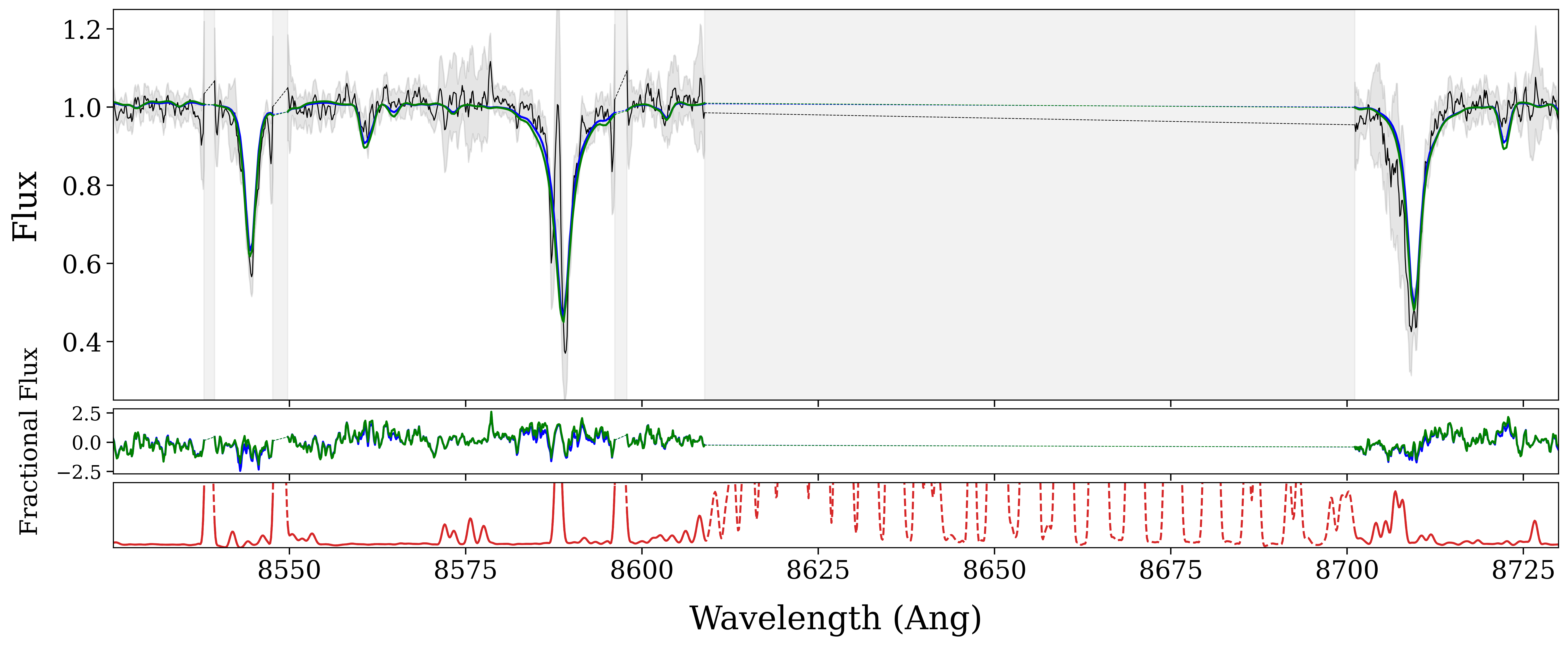}
    \caption{The extracted 1D spectrum of DF9 (\textit{black}), with two fit templates, each with ages and metallicities taken from the central results of past literature (\textit{\citealt{2025ApJ...978...21T} in blue, \citealt{2023MNRAS.524.2624G} in green}), over-plotted. Flux is given in units relative to the continuum, and regions with strong sky line residuals have been masked (\textit{dashed lines}), with the sky spectrum \textcolor{black}(\textit{red}) and residual flux as a fraction of measurement uncertainty plotted in the bottom panels. \label{Fig:Spectrum}}
\end{figure*}

In order to combine the reduced 3D frames, we both convert from vacuum to air wavelengths and apply a constant wavelength shift so that individual frames lie on an equivalent wavelength grid. Next, to spatially align the frames, we collapse the individual reduced frames into 2D by summing along the wavelength direction, excluding wavelengths associated with previously subtracted sky lines (to avoid any remaining residuals), and fit this spatial map of the galaxy with models of DF9 and its brightest star cluster. We then re-center the individual frames, applying shifts such that all frames have a common spatial grid. Masks were created that contain high sigma outliers such as cosmic ray hits; each pixel in the final data cube is the weighted average of all non-masked pixels.


Next, we extract 1D spectra from the combined 3D frame, again fitting 2D models of DF9 and the bright compact objects in the KCWI pointing, including its central star cluster, to a 2D spatial image created by collapsing the 3D frame along the wavelength direction (avoiding sky lines). We then create a 1D spectrum of the galaxy by performing a weighted sum, multiplying the flux at each position by the fitted model of the galaxy. This is done only for pixels where the model of the galaxy is at least 15\% of its peak flux and where the galaxy is a) not contaminated by light from bright compact objects, and b) contributes the majority of flux in the pixel, that is, exceeds the sum of other modeled objects and residuals. All contaminated pixels, in which 2D models of the compact objects' flux is at least 15\% of its own peak flux or exceeds the light of the galaxy in the pixel are excluded. Along with this science spectrum, an error spectrum was generated starting with the square root of the sky counts and following propagation of error rules during reduction.


\subsection{Model Fitting \& MCMC Parameter Estimation} \label{Sec:Fit}

In Figure~\ref{Fig:Spectrum} we show DF9's resulting 1D spectrum in the region of the Ca triplet lines. We fit this spectrum using synthetic stellar population spectra \citep{2009ApJ...699..486C} generated using the MIST isochrones \citep{2016ApJ...823..102C}. Twelve templates with metallicities of $-1.00$, $-1.25$, and $-1.50$ and ages of 7, 8, 9, and 10 Gyr were selected in the broad range expected from past work \citep{2023MNRAS.524.2624G,2025ApJ...978...21T}. In order to construct templates which match the precise resolution of our KCWI observing mode, we first measured the wavelength-dependent instrumental resolution $\sigma_{\rm inst}$ by fitting Gaussians to sky lines. We find an instrumental resolution $\sigma_{\rm inst}  = 0.375$ \AA\:(13.0 km s$^{-1}$) with a standard deviation of 0.007 \AA\:(0.2 km s$^{-1}$). We then verified that the instrumental resolution from sky lines was similar to that determined from arc lamps, and to the expected resolution in the mode that we are using. We then convolved the templates with a kernel to match $\sigma_{\rm inst}$, and verified these final templates had the same resolution as KCWI by fitting the solar spectrum to an equivalent template with the sun's age and metallicity.

Finally, we then fit these templates to the 1D spectrum using the {\tt emcee} Markov chain Monte Carlo (MCMC) sampling algorithm \citep{2013PASP..125..306F} to constrain posterior distributions. These include on the velocity dispersion $\sigma$ as well as a redshift $z = v_{\rm DF9} /c$ (where $v_{\rm DF9}$ is the velocity of the galaxy).  For both the template and the spectrum, we also fit an initial 1st order polynomial, i.e. a line, to the continuum. We ensure burn-in has occurred through visual inspection of walker position distributions, comparing their initial and post–burn-in states to confirm convergence away from the starting conditions (we then discard the initial burn-in phase and use the post–burn-in walker positions to initialize the subsequent production run). We verify the accuracy of our error spectrum by analyzing the distribution of the standardized residuals, which we confirm has a standard deviation of $\approx1$. Uncertainties on fit parameters are determined from 16th and 84th percentiles, with additional uncertainties added in quadrature below. 

We show the result of two such fit template models in Figure~\ref{Fig:Spectrum} (those corresponding to the central values of metallicities and ages from past literature; \citealt{2023MNRAS.524.2624G,2025ApJ...978...21T}). Note that the blue-side of the 8662 \AA\:Calcium triplet line (in rest frame air wavelengths; 8709 \AA\:as observed for DF9) suffers from elevated sky line residuals, with the sky shown in the bottom panel in red, hence the large uncertainties in this region. The fit velocity is similar to the overall average across all templates, $v_{\rm DF9} = 1661\pm2$ km s$^{-1}$. The fit uncertainty includes the quadratic average of MCMC posterior widths, the spread among fits, and a 0.24 km s$^{-1}$ wavelength calibration uncertainty as based on the position offsets of sky lines, all added in quadrature. This updated systemic radial velocity is above the $1644\pm4$ km s$^{-1}$ measured with LRIS from \citet{2025ApJ...988..165K} and is closer to the $1680\pm40$ km s$^{-1}$ prior KCWI result (\citealt{2023MNRAS.524.2624G}; which we have modified based on the suggestion of the author, as \citealt{2025ApJ...988..165K} did previously in their Section 4.6). 

Combining the results of all fits and accounting for the uncertainty in the instrumental resolution, we find an average dispersion of $\sigma_{\rm fit} = 8.7^{+3.4}_{-3.7}$ km s$^{-1}$. 


\section{Results} \label{Sec:Results}


\subsection{Velocity Dispersion} \label{Sec:StellarVelocityDispersion}

The fit velocity dispersion in Section~\ref{Sec:Fit} includes not only the motion of stars in the galaxy's gravitational potential, but also the internal rotation and macroturbulence of the individual stars, as well the rotation of stars in binary systems around their center of mass. For most galaxies, with dynamics dominated by a high dark matter content, these effects are dwarfed by the gravitational motion. But for a system like DF9 with a very low $\sigma_{\rm fit}$, we must take them into account.

For individual stars, rotation and macroturbulence acts to broaden lines. This does not change the star's radial velocity and therefore has no effect on dispersions inferred from the velocities of resolved populations of individual stars as measured for galaxies in the Local Group (as explored in Section~\ref{Sec:ComparisontoWiderPopulation}). However, it does impact dispersions from integrated light as in this work. \citet{2023ApJ...957....6S} estimates the broadening effect to be $\sigma_{\rm broadening} = 5.4\pm2$ km s$^{-1}$ in their analysis of DF4, which has a very similar age and metallicity to DF9 \citep{2019ApJ...874L...5V,2023MNRAS.524.2624G,2025ApJ...978...21T}. See \citet{2023ApJ...957....6S} for a detailed review of how the effect depends on stellar evolution and its estimation in the case of DF4.

For binary systems, orbital motion along the line-of-sight adds a time-varying confusion noise to the star's radial velocity, an effect which depends on the fraction of stars in binaries. This can and does impact dispersions for resolved populations, though in practice observers generally remove binaries rotating along the line-of-sight which appear as clear outliers; the recent revision of the star cluster UNIONS I's dispersion from 3.7 km s$^{-1}$ down to 0.1 km s$^{-1}$ largely due to a single binary \citep{2026ApJ...999L...8C} is an excellent example. However, this piecemeal correction is not possible for dispersions from integrated light as in this work. \citet{2002MNRAS.329..829D} estimates the effect as $\sigma_{\rm binaries} = \alpha \times \sigma_{\rm b}$, where $\alpha$ is the system's binary fraction, 60\% in the solar neighborhood, and $\sigma_{\rm b} = 2.87$ km s$^{-1}$. In general, a system's binary fraction has a strong environmental dependence with dense systems like globular clusters having $\alpha\sim 10\%$ due to dynamical interactions destroying binaries \citep{2025MNRAS.541.2008B} and low stellar density systems as found in Ursa Minor's outskirts having up to $\alpha\sim100\%$ (\citealt{2025arXiv251204477Q}; with a metallicity dependence as well). However, taken as a whole Ursa Minor's $\alpha_{\rm UMi}=0.61^{+0.16}_{-0.20}\%$ binary fraction is nearly identical to that of the solar neighborhood, and given that DF9's surface brightness and metallicity lie about halfway between these two populations and just $\approx1$ mag arcsec$^{-2}$ and $\approx1$ dex apart from each \citep{1998ARA&A..36..435M,2007A&A...462..965M,2018Natur.555..629V,2023NatAs...7..951L,2023MNRAS.524.2624G,2025ApJ...978...21T}, this appears to be an appropriate estimate. We therefore estimate this effect to be $\sigma_{\rm binaries} \approx \alpha_{\rm UMi} \times \sigma_{\rm b} = 1.75^{+0.46}_{-0.57}$ km s$^{-1}$. 

\begin{figure}
    \centering
    \includegraphics[width=0.45\textwidth]{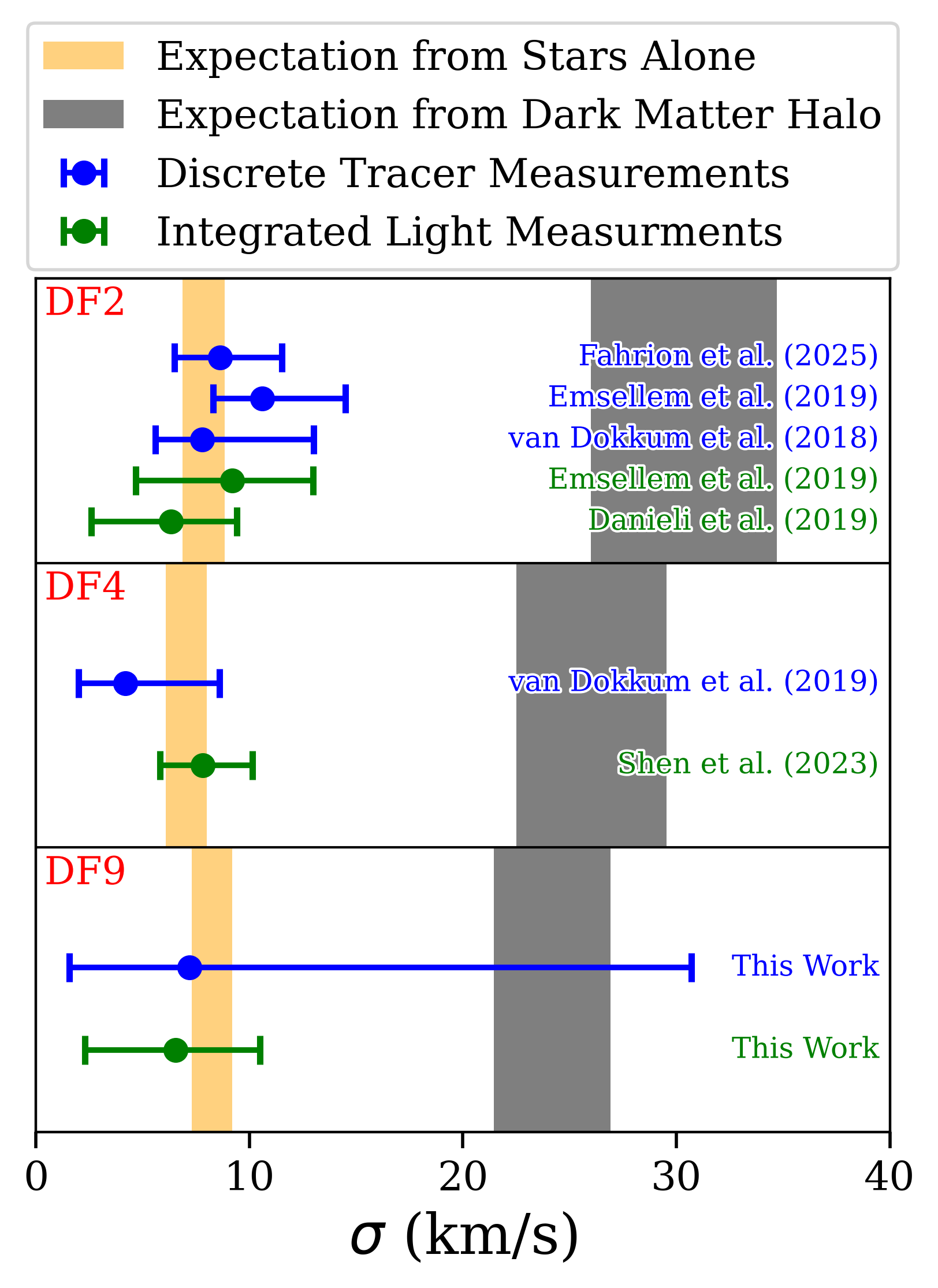}
    \caption{Velocity dispersion as measured from diffuse stellar light (\textit{green}) and from globular cluster and planetary nebulae (\textit{blue}) as compared to that expected for a normal dark matter halo (\textit{black}) and the stars alone (\textit{orange}), for DF2 (\textit{top}), DF4 (\textit{middle}), and -- new in this work -- DF9 (\textit{bottom}). All three galaxies have velocity dispersions consistent with that expected from their stars alone. \label{Fig:Dispersion}}
\end{figure}

Subtracting both contributions in quadrature as \begin{equation}
    \sigma = \sqrt{\sigma_{\rm fit}^2-\sigma_{\rm broadening}^2-\sigma^2_{\rm binaries}}
\end{equation} results in a gravitational dispersion of $\sigma = 6.5^{+3.9}_{-4.3}$ km s$^{-1}$. As shown in Figure~\ref{Fig:Dispersion}, the velocity dispersion of DF9 is consistent with those of DF2 \citep{2018RNAAS...2...54V,2019ApJ...874L..12D,2019A&A...625A..76E,2025A&A...697A.145F} and DF4 \citep{2019ApJ...874L...5V,2023ApJ...957....6S}. We have updated the KCWI integrated light measurement for DF4 to $7.8^{+2.3}_{-2.0}$ km s$^{-1}$, to reflect the impact of binaries, and for DF2 to $6.3^{+3.1}_{-3.7}$ km s$^{-1}$, to reflect both effects described above. We have similarly modified the MUSE integrated light measurement for DF2 to $9.2^{+3.8}_{-4.5}$ km s$^{-1}$ to reflect both effects. The previous measurements from discrete tracers are unaffected by these broadening contributions. 

\subsection{Inferred Mass} \label{Sec:InferredMass}

We may calculate the mass of DF9 within the effective half-light radius $r_{\rm e}$ following \citet{2010MNRAS.406.1220W} as
\begin{equation} \label{Eq:Mass}
    M_{\rm e} = 3 \sigma^2 r_{\rm e, 3D} / G 
\end{equation}
where $r_{\rm e, 3D} \approx 4/3 r_{\rm e}$ is the projected circularized effective half-light radius. Taking the distance to DF9 as 20.6 Mpc, its expected distance based on its position along the trail \citep{2025ApJ...988..165K}, the galaxy's 11.1$^{\arcsec}$ effective radius \citep{2023MNRAS.524.2624G} in physical units is $r_{\rm e} = 1.1$ kpc. As detailed in Appendix~\ref{Sec:Dist}, the precise distance of DF9 has little impact on our findings as $\sigma\propto\sqrt{d}$. Together with our measured dispersion, this implies a dynamical mass within the effective radius of $M_{\rm e} = 0.42^{+0.68}_{-0.37} \times 10^8$ M$_\odot$. 

The dynamical mass is consistent with the stellar mass within the same half-light radius, $M_{\rm e, \star}$. Based on DF9's $m_g = 17.17$ $g$ band apparent magnitude and $M_\star/L_g = 2.2$ $g$ band stellar mass to light ratio \citep{2023MNRAS.524.2624G}, $M_{\rm e, \star} = 0.71\pm0.16 \times 10^8$ M$_\odot$.\footnote{That is, half of DF9's $1.4\times 10^8$ M$_\odot$ total stellar mass. Following \citet{2019ApJ...874L...5V} we assume a $\pm0.5$ uncertainty in $M_\star/L$. Note that the $M_\star/L_g = 2.2$ from \citet{2023MNRAS.524.2624G} is exactly consistent with that found by \citet{2025ApJ...978...21T}.}


\subsection{Comparison to Expected Dispersion} \label{Sec:ComparisontoExpectedDispersion}

The equivalent statement is that the observed velocity dispersion is consistent with that expected from the stars. From Equation~\ref{Eq:Mass}, DF9's $M_{\rm e, \star}$ alone, without accounting for any neutral or ionized gas in the galaxy, would imply a velocity dispersion of $8.3^{+0.9}_{-1.4}$ km s$^{-1}$. This is consistent with the observed value, as shown in Figure~\ref{Fig:Dispersion}, as was similarly true for DF2 and DF4. Note in Figure~\ref{Fig:Dispersion} we have calculated DF2 and DF4's expected dispersion based on updated distance measurements \citep{2020ApJ...895L...4D,2021ApJ...914L..12S}, though these results also are insensitive to distance, as demonstrated in Appendix~\ref{Sec:Dist}.

We can also compare the observed velocity dispersion to the expected value if DF9 had a ``standard'' dark matter halo. For an NFW halo \citealt{1997ApJ...490..493N} following the stellar mass-halo mass relation, one would expect DF9 to have a $M_{\rm h} = 1.4\pm0.7 \times 10^{10}$ M$_\odot$ dark matter halo \citep{2023MNRAS.522.2696Z,2023MNRAS.519..871Z} and a dispersion of $24 \pm 3$ km s$^{-1}$. To calculate this we use the same WMAP9 flat $\Lambda$CDM cosmology \citep{2013ApJS..208...19H} and halo-concentration relation \citep{2007MNRAS.378...55M} adopted by \citet{2023MNRAS.519..871Z}, then evaluate the halo mass at $r_{\rm e}$ to find $\sigma$ from Equation~\ref{Eq:Mass}. Uncertainties are estimated from the reported 0.3 dex scatter. We also explored multiple other stellar mass-halo mass relations from literature, such as that from \citet{2023ApJ...956....6D} which yielded a larger expected dispersion of 32 km s$^{-1}$. As noted in \citet{2023MNRAS.519..871Z}, their halo mass estimates are significantly lower for a given stellar mass than in prior literature (see their Fig. 4 comparison to \citealt{2010ApJ...717..379B,2019MNRAS.488.3143B}). Thus, the $24 \pm 3$ km s$^{-1}$ figure based on their work may be taken as a conservative, low estimate compared to the $>30$ km s$^{-1}$ expected from literature more broadly.

In Figure~\ref{Fig:Dispersion}, we see that, as for DF2 and DF4, DF9's expected dispersion for a standard dark matter halo is inconsistent with the observed dispersions. This discrepancy would be even larger for the stellar mass-halo mass relation from \citet{2023ApJ...956....6D}. Note the expected halo mass for DF2 and DF4 is calculated in the same manner as for DF9 described above.


\subsection{Constraints from Star Clusters} \label{Sec:GlobularClusterVelocityDispersion}

While our main result is attributable solely to deep, high-resolution spectra for stellar kinematics, we note that \citet{2025ApJ...988..165K} measured the velocities of two star clusters in DF9 from Keck/LRIS spectra. One of these is a likely globular cluster; the other is a bright central star cluster. With two tracers of the potential the uncertainty in the kinematics is obviously very high;  however, they can provide a useful consistency check. The argument is fully detailed in Appendix~\ref{Sec:LowN}. In brief, if the two clusters have a large velocity difference, it would rule out models without dark matter. For example, for $\Delta v>$20 km s$^{-1}$ for both clusters (with the zeropoint defined by the diffuse light), a velocity dispersion of $\sigma = 7$ km s$^{-1}$ could be ruled out as it would require that 2 out of 2 samples are 3$\sigma$ outliers, an extreme statistical anomaly. This argument is essentially the same as that used by \citet{1983ApJ...266L..11A}, where velocities of 3 stars in the Draco dwarf spheroidal were used to argue that the galaxy contained dark matter. Here, we use the same argument with the velocities of 2 star clusters relative to the diffuse stellar light of the galaxy. We emphasize the reverse is \textit{not} true, i.e., we cannot rule out the presence of dark matter this way, given the long tails in the inferred uncertainties in Appendix~\ref{Sec:LowN}.

In order to calculate the constraint provided by the clusters we construct the likelihood function as
\begin{equation} \label{Eq:Liklihood}
\mathcal{L}\left(\sigma\right)
= \prod_{i=1}^{i \le N}
\frac{e^{-\frac{1}{2} \left(\frac{v_i - \mu}{\sigma^2 + \delta v^2_i + \delta \mu^2}\right)^2}}{\sqrt{2\pi \left(\sigma^2 + \delta v^2_i + \delta \mu^2\right)}},
\end{equation}
where $v_i$, $\delta v_i$ are the velocities of the clusters and their associated uncertainties and $\mu$, $\delta \mu$ are the  systemic velocity of the galaxy and its uncertainty. In doing so, we incorporate the velocity of the diffuse light information by assuming that it is the true mean systematic velocity of the system $\mu$. Given the asymmetric uncertainties in \citet{2025ApJ...988..165K}, for both $v_i$ and $\mu$ we take $\delta x$ to be $\frac{1}{2}\left(\delta x_+ + \delta x_-\right)$. As in \citet{2023ApJ...951...77T}, for such a low number of tracers, our results depend on the prior utilized; in particular, a flat prior is systematically biased to produce overestimates, whereas a Jeffreys prior better recovers the true underlying velocity dispersions. We therefore apply the prior correction term given in Equation 2 of \citet{2023ApJ...951...77T}.

In Figure~\ref{Fig:Dispersion} we plot the LRIS cluster constraint, 7$^{+23}_{-6}$ km s$^{-1}$ (68\% confidence), alongside the velocity dispersion from the stellar kinematics. We find that the fact that both clusters have very similar radial velocities as the diffuse light ($1649\pm2$ km s$^{-1}$ and $1645^{+5}_{-6}$ km s$^{-1}$ as compared to $1644\pm4$ km s$^{-1}$; \citealt{2025ApJ...988..165K}) is consistent with the low dispersion that was measured with KCWI.


\subsection{Comparison to Wider Population} \label{Sec:ComparisontoWiderPopulation}

\begin{figure*}
    \centering
    \includegraphics[width=0.52763874949\textwidth]{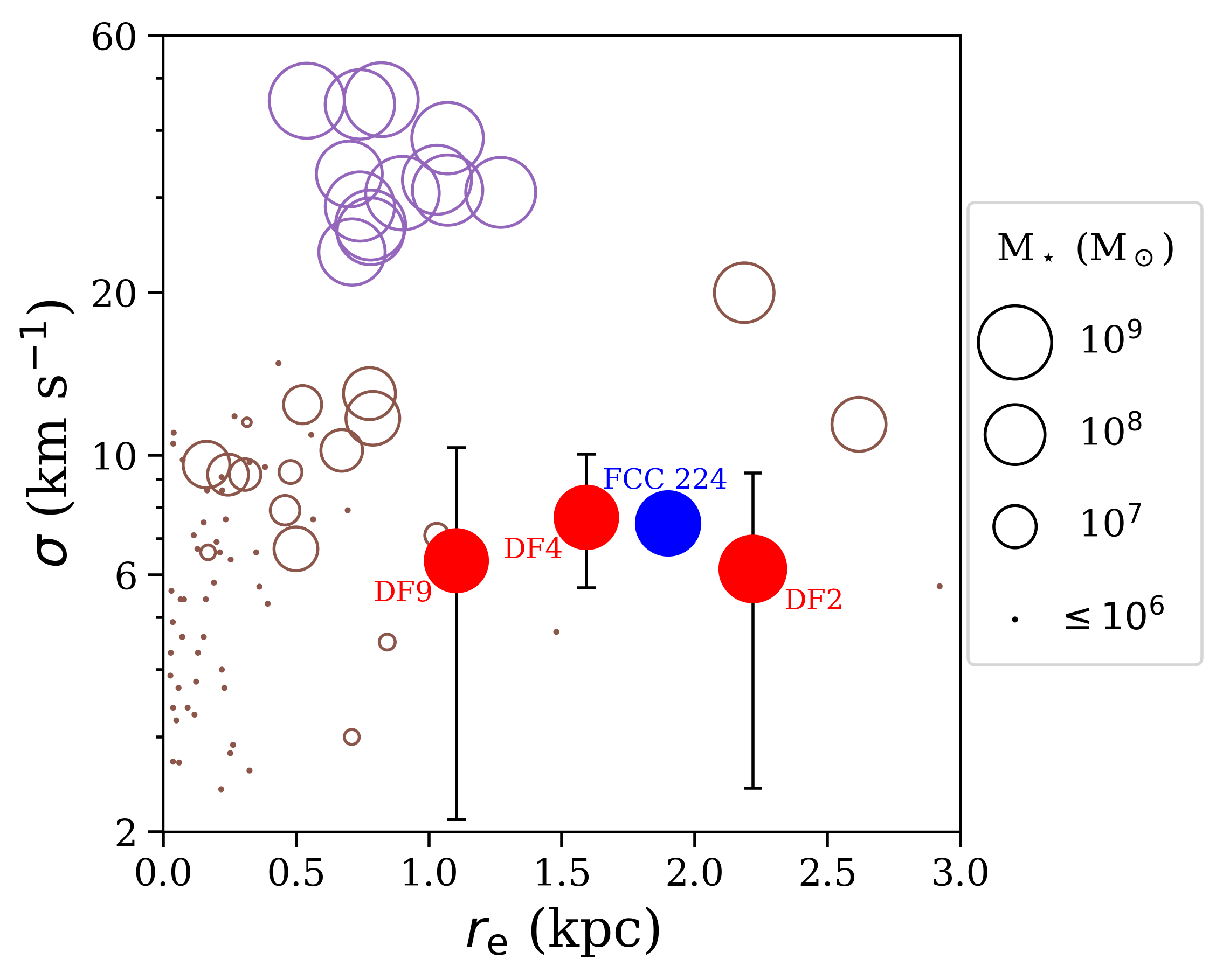}\quad\includegraphics[width=0.4423612505\textwidth]{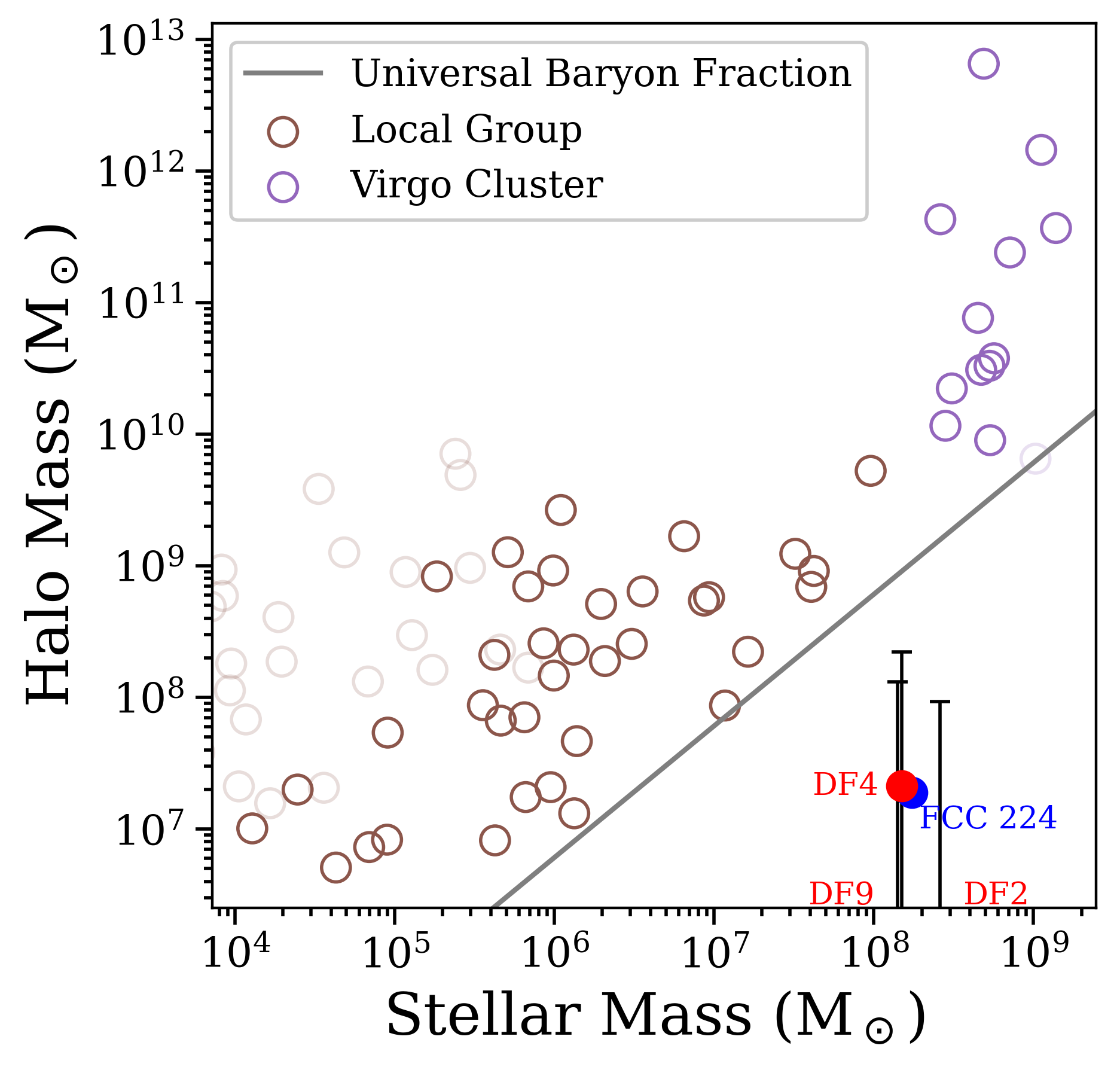} 
    \caption{A comparison between DF2, DF4, and DF9 (\textit{filled red circles}) and a wider population of Local Group (\textit{open brown circles}) and Virgo Cluster dwarf galaxies (\textit{open purple circles}), both in terms of measured velocity dispersion versus size (\textit{left, with stellar mass indicated by markersize in logscale}), and inferred halo mass versus stellar mass (\textit{right, with unreliable halo mass estimates given in light shading}). FCC 224, a similar galaxy identified by \citet{2025A&A...695A.124B}, is also shown (\textit{filled blue circle}). Galaxies with similar velocity dispersions to DF2, DF4, and DF9 tend to have $>$100$\times$ lower stellar masses and 2$-$6$\times$ smaller radii, while galaxies with similar stellar masses tend to have $>$100$\times$ more massive dark matter halos. Even at the upper-bounds of their uncertainties, DF2, DF4, and DF9 fall below the universal baryon fraction (\textit{gray line}); galaxies that formed within dark matter halos should not lie in this region, even if they convert all their baryons into stars.\label{Fig:Pop}}
\end{figure*}

To better visualize the unusual nature of the galaxies, in the left panel of Figure~\ref{Fig:Pop} we compare the observed sizes and velocity dispersions of DF2, DF4, and DF9 (as measured with KCWI) to a wider population of galaxies. Our comparison includes both galaxies in the Local Group (\citealt{2012AJ....144....4M}; using the most up-to-date \href{https://www.cadc-ccda.hia-iha.nrc-cnrc.gc.ca/en/community/nearby/}{Jan2021 catalog version}) as well as galaxies in the Virgo Cluster \citep{2003AJ....126.1794G}. In order to only consider directly comparable dwarf galaxy velocity dispersions, we restrict the sample to dispersion-supported, non-rotating dwarfs that satisfy the $v_{rot}/\sigma$ criteria of \citet{2003AJ....126.1794G}. Note that while the \citet{2003AJ....126.1794G} sample is measured in integrated light, all dispersions are too large to be noticeably affected by the stellar motion discussed in Section~\ref{Sec:StellarVelocityDispersion}. Datapoints are given sizes which scale with the base 10 logarithm of their stellar mass (as in Extended Data Figure 2 of \citealt{2018Natur.555..629V}).

We find that DF2, DF4, and DF9 are all outliers in the $r_{\rm e} - \sigma - M_\star$ plane. Galaxies with similar velocity dispersions are several times smaller in effective half-light radius and have several orders of magnitude lower stellar mass, whereas galaxies with similar stellar masses have much larger velocity dispersions. 

In order to understand the implications for the galaxies' dark matter content, in the right panel of Figure~\ref{Fig:Pop} we show inferred halo masses using the same sample. We follow the same procedure as \citet{2023MNRAS.519..871Z} in order to calculate halo mass: We first calculate $M_{\rm e}$ from Equation~\ref{Eq:Mass}, using the observed properties of the galaxies. We then subtract off the stellar mass $M_{\rm e, \star}$ and run a least squares fit to find the halo mass which has a mass of $M_{\rm e, DM} = M_{\rm e}-M_{\rm e, \star}$ at $r_{\rm e}$, using NFW potentials with the same cosmology and concentration relation as \citet{2023MNRAS.519..871Z}. Again following the procedure of \citet{2023MNRAS.519..871Z}, we `add back' the baryons by multiplying by $(1+\Omega_{\rm b}/\Omega_{\rm DM}) = (1+0.0464/0.235) \approx1.197$ such that $M_{\rm h}$ represents the overall halo mass, and mark results for objects that fall below the 10 pc effective radius limit or above the $M_{\rm e, DM}/M_{\rm e} = 0.99$ limit of \citet{2023MNRAS.519..871Z} as unreliable (indicated in light shading). For both sample's $V$ band magnitudes, we assume a global $M_\star/L_V = 2$ as found from stellar-dominated globular cluster dynamics \citep{2005ApJS..161..304M}.

As has been long established, low mass dwarf galaxies typically have very high dark matter fractions, and indeed in Figure~\ref{Fig:Pop} most galaxies in our sample have inferred halo masses which are two orders of magnitude greater than their stellar mass \citep{2023MNRAS.519..871Z,2013ApJ...770...57B}. The line indicates the universal baryon fraction; no galaxies are expected below this line unless their dark matter has been affected due to interactions like those discussed in Section~\ref{Sec:FormationTheories}. As expected, dwarfs in the Local Group and Virgo samples all fall above the universal baryonic fraction, as they can at most convert all the baryons in their halo into stars and no more than that.\footnote{The only dwarf from our comparison sample to fall on the universal baryon fraction is VCC 917. Notably, though followup work on the sample has generally been consistent within a few km s$^{-1}$, a clear exception is VCC 917 which reaches $\sigma > 40$ km s$^{-1}$ at $r_{\rm e}$ \citep{2009MNRAS.394.1229C}. While we maintain the original $\sigma$ for consistency, we give it light shading to indicate that the updated measurement would place its halo at $>10^{11}$ M$_\odot$, i.e. a two orders of magnitude greater than its stellar mass and far above the universal baryon fraction.} 

However,  DF2, DF4, and DF9 are again clear outliers. While their masses are consistent with an extremely small `normal' dark matter halo, such a halo would lie below the universal baryonic fraction. This is unphysical in the standard model for galaxy formation, as it would imply that the galaxies started with an order of magnitude more baryons than the rest of the universe (and converted all these baryons into stars).  The more straightforward interpretation is that the galaxies lack dark matter due to a unique formation mechanism. We note that DF2, DF4 and DF9 are not the only galaxies falling in this regime; in particular, \citet{2025A&A...695A.124B} recently identified a similar system, FCC 224 (see Figure~\ref{Fig:Pop}), which may have formed in a similar way as the NGC 1052 trail galaxies.


\section{Discussion} \label{Sec:Discussion}


\subsection{A Third Galaxy without Dark Matter} \label{Sec:AThirdGalaxywithoutDarkMatter}

If DF2, DF4, and DF9 were the only galaxies known, one would not conclude there is any need for dark matter. For all three galaxies, their dynamical mass as inferred from the gravitational motion of their stars is  consistent with their stellar mass as estimated from the light they emit. While within measurement uncertainties their dynamical masses are consistent with a small amount of mass beyond that estimated photometrically from their stars, the implied baryon fractions would be multiple times larger than the universal baryon fraction. Though DF2 has little gas \citep{2019ApJ...871L..31S,2019MNRAS.482L..99C}, it is possible that DF9's baryonic mass, including both stars and gas, accounts for an even larger portion of its mass, leaving even less room for dark matter in the galaxy.

While in Section~\ref{Sec:ComparisontoExpectedDispersion} we considered a ``standard'' NFW halo, \citet{2024A&A...689A.161A} also considers the case of a cored dark matter halo for DF2, and finds that such a halo would only be consistent with a normal amount of dark matter if it had an extreme scale radius and cutoff radius.\footnote{Buzzo et al., in private communication regarding a submitted paper, finds the same while reviewing both DF2 and DF4 along with a system of galaxies bearing close similarities to the NGC 1052 trail.} This is not physically motivated, whereas the mass of the stars alone naturally explains the trail galaxies' dynamics. If the trail galaxies do all have such unusual and extreme cored halos, the ultimate conclusion is the same as in the case that they entirely lack dark matter; i.e. that they formed together through an extreme galaxy formation channel.


\subsection{Formation Theories} \label{Sec:FormationTheories}

While almost every other dwarf galaxy previously studied is strongly dark matter dominated,  it is remarkable to find DF2, DF4, and now DF9 all seemingly without dark matter and part of the same tight, kinematically-connected line of galaxies. It is all but certain that this trail of galaxies formed together through a mechanism that caused their dark matter deficiency.

DF9 lacking dark matter was a specific and original prediction of the `bullet dwarf' high-speed collision model, as discussed in Section~\ref{Sec:Introduction}. Nevertheless, \citet{2025ApJ...988..165K} includes an in-depth review of potential other formation theories and their likelihood for the observed galaxy trail. These include the high-speed bullet dwarf collision scenario discussed in Section~\ref{Sec:Introduction}, as well as gas flung out by quasars \citep{1998MNRAS.298..577N}, tidal dwarfs formed along tidal features created by a merger of massive galaxies (see e.g. \citealt{2019A&A...628A..60F}), and tidal stripping of an infalling stream of galaxies (similar to the linear substructures seen by \citealt{2023Natur.623..296W}). Such alternative models generally involve `top-down' formation with baryons being separated out from their original dark matter halo due to interactions with other galaxies, and therefore all would similarly expect DF9 to lack dark matter. Notably, the NGC 1052 trail before its discovery was only predicted by the high-speed collision model \citep{2020ApJ...899...25S}.



\subsection{Future Efforts} \label{Sec:FutureEfforts}

It would be interesting to extend these measurements to the remaining trail galaxies, but this will be very difficult. With the DF9 result, we have now measured dispersions for all the $>10^8$ M$_\odot$ trail galaxies; the remaining trail galaxies have stellar masses ranging from $4\times10^6$ M$_\odot$ to $3\times10^7$ M$_\odot$ and dispersions of just $1-4$ km s$^{-1}$. Another avenue to learn more about the system is to measure the spatial distribution and kinematics of neutral and ionized gas in the group. If any remaining trail galaxies have gas, this could provide dynamical constraints for objects that are too faint for optical spectroscopy. Such data could also show whether any gas left behind by the original bullet dwarf collision coincident with the trail is still visible. This would be `smoking gun' evidence for such an event. Moreover, a constraint on the gas content of the three galaxies with measured kinematics would better constrain their  total baryonic masses, as discussed in Section~\ref{Sec:AThirdGalaxywithoutDarkMatter}, and potentially even help distinguish between dark matter models \citep{2025arXiv250924270W}. 


\section{Summary and Conclusion} \label{Sec:Summary and Conclusion}

In this work we measured the dynamical mass of DF9, a galaxy on a kinematically connected linear trail which includes the `galaxies lacking dark matter' DF2 and DF4, finding that DF9 also lacks dark matter. Unlike the vast majority of other dwarf galaxies, the objects on this unique trail appear to be dark matter-free. One may naturally conclude that the galaxies formed together in an atypical way that caused their dark matter deficiency.

DF9's lack of dark matter was a strong, potentially falsifying prediction of \citet{2022Natur.605..435V} after discovering the trail and postulating that it formed in a high-speed bullet dwarf collision (as first described for DF2 alone by \citealt{2019MNRAS.488L..24S}), and DF9 would not be an object of interest at all were it not for the bullet dwarf theory. Nevertheless, however it formed, this third galaxy lacking dark matter, and the rest of the trail, represents the most extreme end of galaxy formation, and is key to understanding the nature of dark matter.


\section*{Acknowledgments} \label{Sec:Acknowledgments}
We thank W.\ Cerny for helpful conversations regarding the impact of binary motion on inferred velocity dispersion measurements. Support from STScI grants HST-GO-14644, HST-GO-15695, and HST-GO-15851, and HST-GO-16912 is gratefully acknowledged. Some of the data presented herein were obtained at Keck Observatory, which is a private 501(c)3 non-profit organization operated as a scientific partnership among the California Institute of Technology, the University of California, and the National Aeronautics and Space Administration. The Observatory was made possible by the generous financial support of the W. M. Keck Foundation. This research is based on observations made with the NASA/ESA Hubble Space Telescope obtained from the Space Telescope Science Institute, which is operated by the Association of Universities for Research in Astronomy, Inc., under NASA contract NAS 5–26555. These observations are associated with programs 14644, 15695, 15851, and 16912. 

\vspace{5mm}

\textit{HST} data utilized in this paper can be accessed from the Mikulski Archive for Space Telescopes (MAST) at the Space Telescope Science Institute via \dataset[10.17909/hj6b-7268]{https://doi.org/10.17909/hj6b-7268}.

\facilities{Keck (KCWI), Keck (LRIS), and \textit{HST} (ACS).}

\software{KCWI DRP and astropy \citep{2013A&A...558A..33A,2018AJ....156..123A}.}


\appendix

\section{The Utility of Low N Consistency Checks} \label{Sec:LowN}

\begin{figure}
    \centering
    \includegraphics[width=0.49\textwidth]{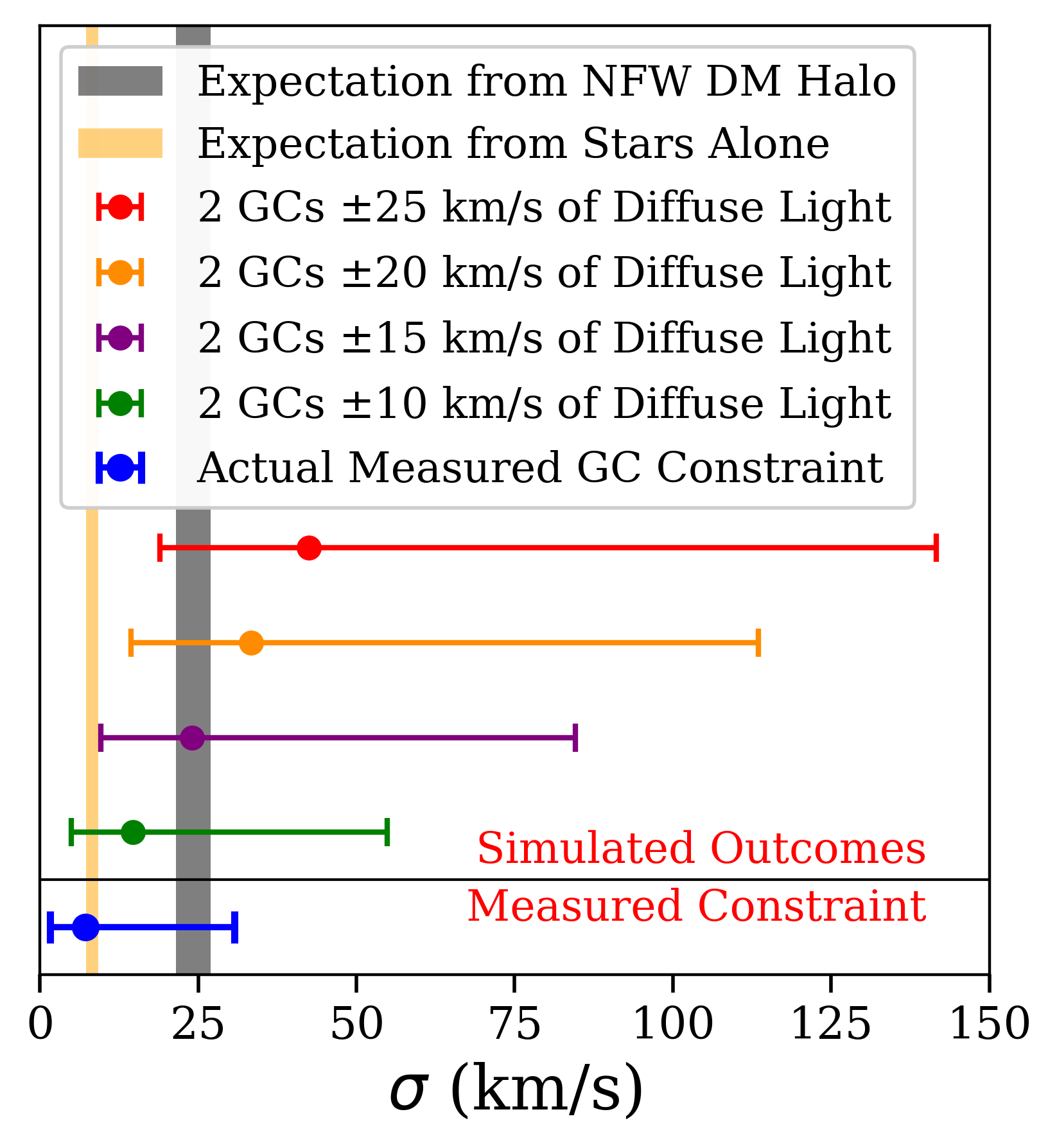} 
    \caption{A demonstration of how the radial velocities of two globular clusters alone can indicate the presence of dark matter in a galaxy in the same manner that \citet{1983ApJ...266L..11A} used to show the Draco dwarf spheroidal contains dark matter. This is markedly not the case for DF9. \label{Fig:GC}}
\end{figure}

In the main text we point out that, although we only have velocity measurements for two of DF9's star clusters, the difference between the presence or entire absence of a massive dark matter halo is so great that a meaningful consistency check can nevertheless be obtained. This argument mirrors the work done by \citet{1983ApJ...266L..11A}, who used just three stars in the Draco dwarf spheroidal to successfully argue that the galaxy contained dark matter. To demonstrate how this can be done, in Figure~\ref{Fig:GC} we artificially alter the radial velocities for the two clusters used in the main text such that they are 10, 15, 20, and 25 km s$^{-1}$ above and below the diffuse light of the galaxy. For instance, the `$\pm10$ km s$^{-1}$' datapoint involves placing one cluster at $v_{\rm DF9}+10$ km s$^{-1}$ and one at $v_{\rm DF9}-10$ km s$^{-1}$, and repeating the procedure from Section~\ref{Sec:GlobularClusterVelocityDispersion} to derive the inferred constraint. We retain the uncertainties from the original measurements as in Section~\ref{Sec:GlobularClusterVelocityDispersion}.

In Figure~\ref{Fig:GC} we see that two objects $\pm$20 km s$^{-1}$ apart from the galaxy's radial velocity can theoretically demonstrate that the mass of the galaxy far exceeds that of its stars alone and would require dark matter to explain its kinematics. However, this is markedly \textit{not} the case for DF9.

\section{The Low Distance Sensitivity of our Results} \label{Sec:Dist}

In the main text we assume that DF9 is at 20.6 Mpc, an estimate based on the positions and distances of DF2 ($21.7 \pm 1.2$ Mpc; \citealt{2021ApJ...914L..12S,2023ApJ...957....6S}) and DF4 ($20.0 \pm 1.6$ Mpc; \citealt{2020ApJ...895L...4D}) under the assumption that the trail galaxies follow a linear relation between their position along their trail and their distances \citep{2025ApJ...988..165K}. Given the Mpc-scale uncertainties in DF2 and DF4's best distance estimates, and the modest scatter in the linear distance expectation as seen in simulation \citep{2024ApJ...966...72L}, it is worth considering how much our result depends on the assumed distance of DF9.

\begin{figure}
    \centering
    \includegraphics[width=0.49\textwidth]{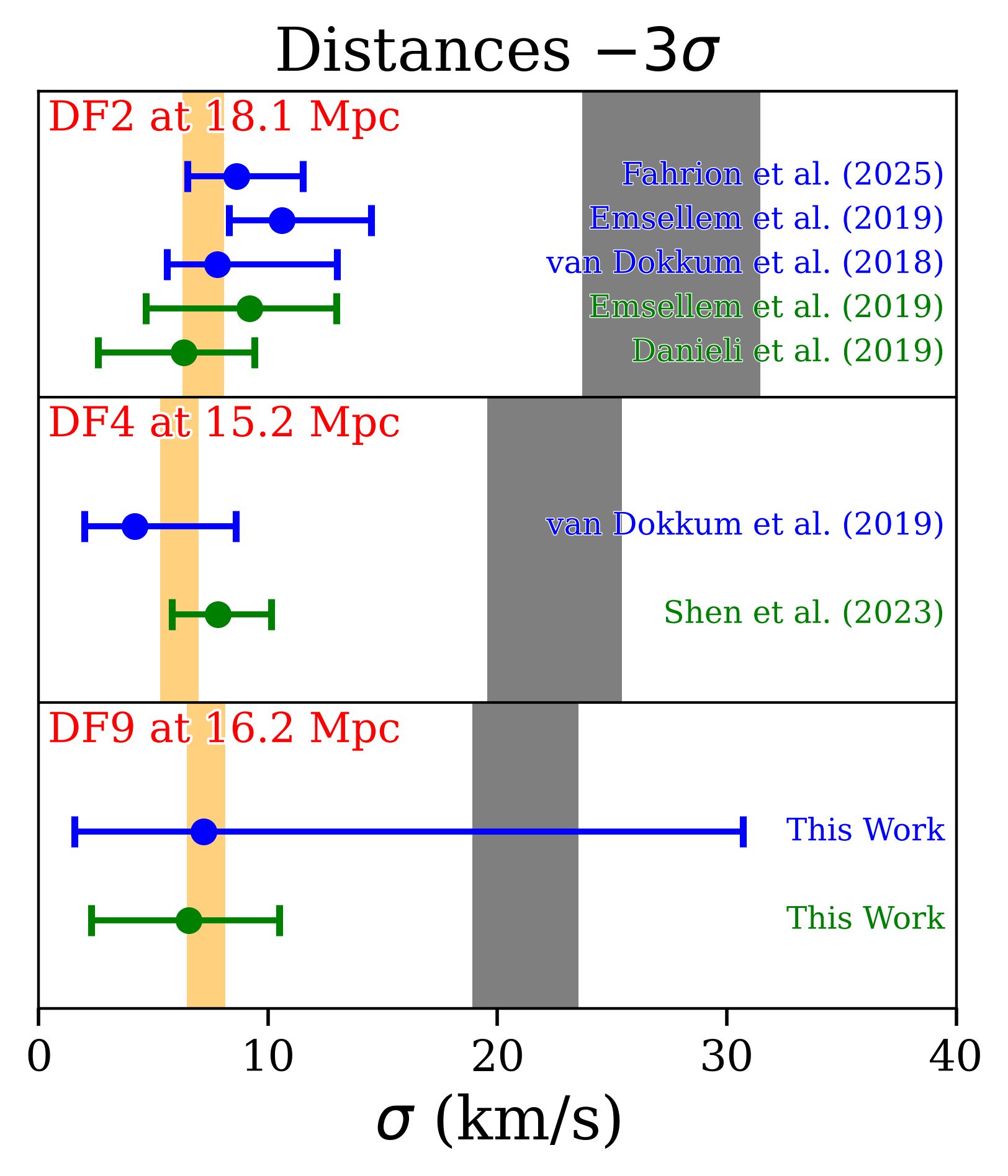}\includegraphics[width=0.49\textwidth]{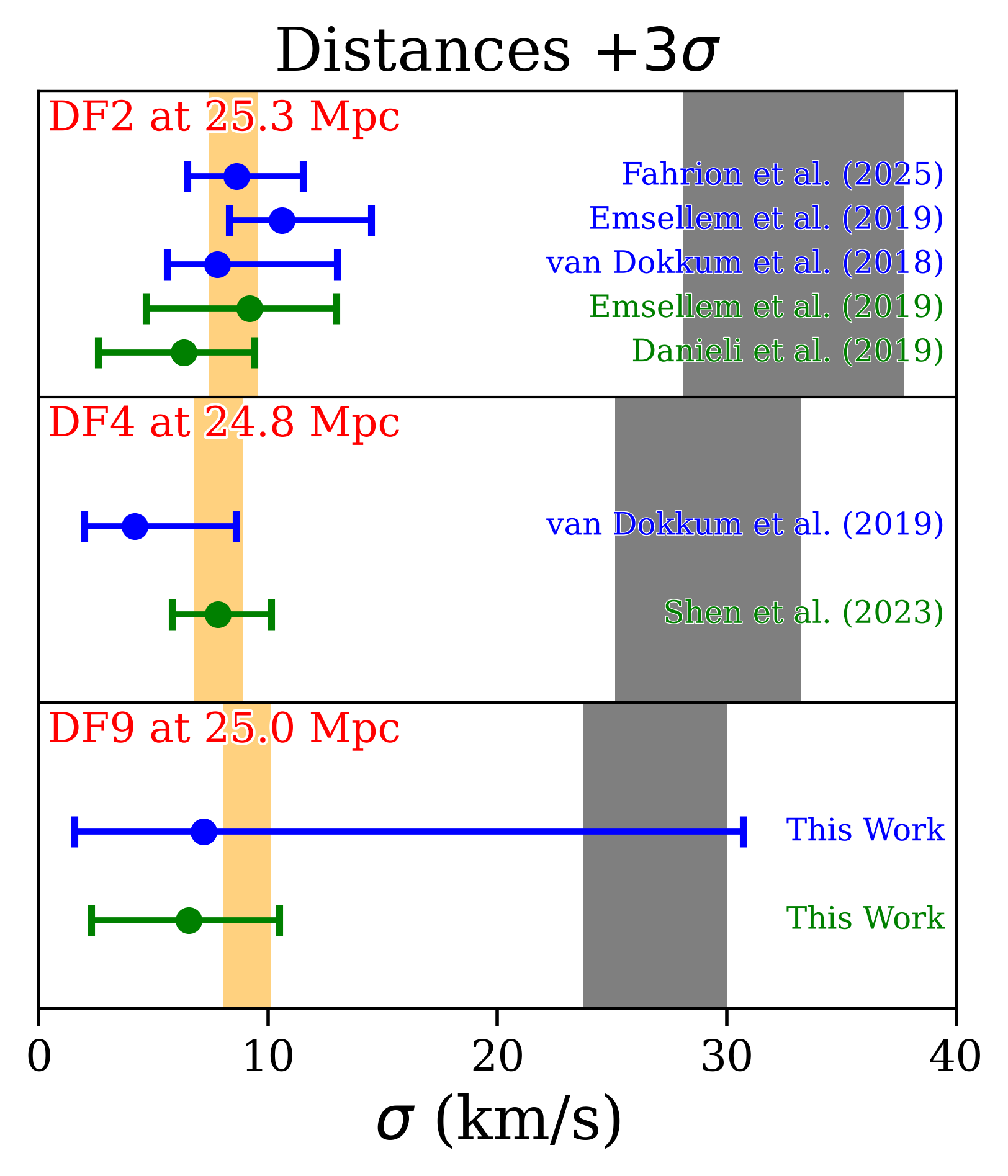} 
    \caption{A demonstration of how the assumed distance has little effect on our findings as originally shown in Figure~\ref{Fig:Dispersion}. We show the effect of shifting the distance of DF9 down to 16.2 Mpc (\textit{left panel}) and up by 25.0 Mpc (\textit{right panel}) on the dispersion expected for a normal dark matter halo (\textit{black}) and the stars alone (\textit{orange}). The same is shown for DF2 and DF4. As in Figure~\ref{Fig:Dispersion}, we give the dispersion as measured both from diffuse stellar light (\textit{green}) and globular clusters (\textit{blue}). \label{Fig:Distance}}
\end{figure}

In Figure~\ref{Fig:Distance} we show the impact of placing DF9 at 16.2 Mpc (i.e. the implied distance if DF2 and DF4 were both shifted $3\sigma_{\rm d}$ down to 18.1 Mpc and 15.2 Mpc, respectively) and 25.0 Mpc (i.e. the implied distance if DF2 and DF4 were each shifted up by $3\sigma_{d}$ to 25.3 and 24.8 Mpc, respectively). We find that it has little effect on our results. This is because the expected dispersion due to the stars alone is proportional to $\sqrt{M/r} \propto \sqrt{d}$, meaning that this 21\% difference in distance affects the expected dispersion by only 11\%.

Note that all elements of Figure~\ref{Fig:Distance}, for DF2, DF4, and DF9, are recalculated based on the adjusted distances, including the stellar mass enclosed and its expected dispersion as well as the dark matter halo potential and its expected dispersion.

\bibliographystyle{aasjournal}
\bibliography{bibliography}

\end{document}